# An Information Theoretic Representation of Agent Dynamics as Set Intersections

Samuel Epstein and Margrit Betke

Boston University, 111 Cummington St, Boston, MA 02215
{samepst,betke}@cs.bu.edu

**Abstract.** We represent agents as sets of strings. Each string encodes a potential interaction with another agent or environment. We represent the total set of dynamics between two agents as the intersection of their respective strings, we prove complexity properties of player interactions using Algorithmic Information Theory. We show how the proposed construction is compatible with Universal Artificial Intelligence, in that the AIXI model can be seen as universal with respect to interaction.[1]

## 1 Introduction

Whereas classical Information Theory is concerned with quantifying the expected number of bits needed for communication, Algorithmic Information Theory (AIT) principally studies the complexity of individual strings. A central measure of AIT is the Kolmogorov Complexity $C(x)$ of a string $x$, which is the size of the smallest program that will output $x$ on a universal Turing machine. Another central definition of AIT is the universal prior $\mathbf{m}(x)$ that weights a hypothesis (string) by the complexity of the programs that produce it [LV08]. This universal prior has many remarkable properties; if $\mathbf{m}(x)$ is used for induction, then any computable sequence can be learned with only the minimum amount of data. Unfortunately, $C(x)$ and $\mathbf{m}(x)$ are not finitely computable. Algorithmic Information Theory can be interpreted as a generalization of classicial Information Theory [CT91] and the Minimum Description Length principal. Some other applications include universal PAC learning and Algorithmic Statistics [LV08,GTV01].

The question of whether AIT can be used to form the foundation of Artificial Intelligence was answered in the affirmative with Hutter's Universal Artificial Intelligence (UAI) [Hut04]. This was achieved by the application of the universal prior $\mathbf{m}(x)$ to the cybernetic agent model, where an agent communicates with an environment through sequential cycles of action, perception, and reward. It was shown that there exists a universal agent, the AIXI model, that inherits many universality properties from $\mathbf{m}(x)$. In particular, the AIXI model will converge to achieve optimal rewards given long enough time in the environment.

---

[1] The authors are grateful to Leonid Levin for insightful discussions and acknowledge partial support by NSF grant 0713229.

As almost all AI problems can be formalized in the cybernetic agent model, the AIXI model is a complete theoretical solution to the field of Artificial General Intelligence [GP07].

In this paper, we represent agents as sets of strings and the potential dynamics between them as the intersection of their respective sets of strings (Sec. 2). We connect this interpretation of interacting agents to the cybernetic agent model (Sec. 2.2). We provide background on Algorithmic Information Theory (Sec. 3) and show how agent learning can be described with algorithmic complexity (Sec. 4). We apply combinatorial and algorithmic proof techniques [VV10] to study the dynamics between agents (Sec. 5). In particular, we describe the approximation of agents (Th. 2), the conditions for removal of superfluous information in the encoding of an agent (Th. 3), and the consequences of having multiple payers achieving the same rewards in an environment (Th. 4). We show how the interpretation given in Sec. 2 is compatible with Universal Artificial Intelligence, in that the AIXI model has universality properties with respect to our definition of "interaction" (Sec. 6).

## 2 Interaction as Intersection

We define players $A$ and $B$ as two sets containing strings of size $n$. Each string $x$ in the intersection set $A \cap B$ represents a particular "interaction" between players $A$ and $B$. We will use the terms *string* and *interaction* interchangeably. This set representation can be used to encode non-cooperative games (Sec. 2.1) and instances of the cybernetic agent model (Sec. 2.2). Uncertainties in instances of both domains can be encoded into the size of the intersections. The amount of uncertainty between players is equal to $|A \cap B|$. If the interaction between the players is deterministic then $|A \cap B| = 1$. If uncertainty exists, then multiple interactions are possible and $|A \cap B| > 1$. We say that player $A$ *interacts* with $B$ if $|A \cap B| > 0$.

### 2.1 Non-cooperative Games

Sets can be used to encode adversaries in sequential games [RN09], where agents exchange a series of actions over a finite number of plies. Each game or interaction consists of the recording of actions by adversaries $\alpha$ and $\beta$, with $x = (a_1, b_1)(a_2, b_2)(a_3, b_3)$ for a game of three rounds. The player (set) representation $A$ of adversary $\alpha$ is the set of games representing all possible actions by $\alpha$'s adversary with $\alpha$'s responding actions, and similarly for player $B$ representing adversary $\beta$. An example game is rock-paper-scissors where adversaries $\alpha$ and $\beta$ play two sequential rounds with an action space of $\{R, P, S\}$. Adversary $\alpha$ only plays rock, whereas adversary $\beta$ first plays paper, then copies his adversary's play of the first round. The corresponding players (sets) $A$ and $B$ can be seen in Fig. 1a. The intersection set of $A$ and $B$ contains the single interaction $x =$ "$(R, P)(R, R)$," which is the only possible game (interaction) that $\alpha$ and $\beta$ can play.

*Example 1 (Chess Game).* We use the example of a chess game with uncertainty between two players: Anatoly as white and Boris as black. An interaction $x \in$ $A \cap B$ between Anatoly and Boris is a game of chess played for at most $m$ plies for each player, with $x = a_1 b_1 a_2 b_2 \ldots a_m b_m = ab_{1:m}$. The chess move space $\mathcal{V} \subset \{0,1\}^*$ has a short binary encoding, whose precise definition is not important. If the game has not ended after $m$ rounds, then the game is considered a draw. Both players are nondeterministic, where at every ply, they can choose from a selection of moves. Anatoly's decisions can be represented by a function $f_A : \mathcal{V}^* \to 2^{\mathcal{V}}$ and similarly Boris' decisions by $f_B$. Anatoly can be represented by a set $A$, with $A = \{ab_{1:m} : \forall_{1 \leq k \leq m}\ a_k \in f_A(ab_{1:k-1})\}$, and similarly Boris by set $B$. Their intersection, $A \cap B$, represents the set of possible games that Anatoly and Boris can play together.

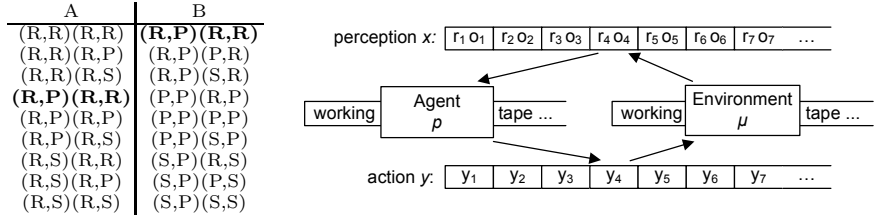

**Fig. 1.** (a) The set representation of players $A$ and $B$ playing two games of paper, rock, scissors. The intersection set of $A$ and $B$ contains the single interaction $x = $ "$(R,P)(R,R)$." (b) The cybernetic agent model.

Generally, sets can encode adversaries of non-cooperative normal form games, with their interactions consisting of pure Nash equilibriums [RN09]. A normal form game is defined as $(p, q)$ with the adversaries represented by normalized payoff functions $p$ and $q$ of the form $\{0,1\}^n \times \{0,1\}^n \to [0,1]$. The set of pure Nash equilibriums is $\{\langle x, y\rangle : p(x,y) = q(y,x) = 1\}$. For each payoff function $p$ there is a player $A = \{\langle x, y\rangle \mid p(x,y) = 1\}$, and for each payoff function $q$ there is player $B$. The intersection of $A$ and $B$ is equal to the set of pure Nash equilibriums of $p$ and $q$.

### 2.2 Cybernetic Agent Model

The interpretation of "interaction as intersection" is also applicable to the cybernetic agent model used in Universal Artificial Intelligence [Hut04]. With the cybernetic agent model, there is an agent and an environment communicating in a series of cycles $k = 1, 2, \ldots$ (Fig. 1b). At cycle $k$, the agent performs an action $y_k \in \mathcal{Y}$, dependent on the previous history $yx_{<k} = y_1 x_1 \ldots y_{k-1} x_{k-1}$. The environment accepts the action and in turn outputs $x_k \in \mathcal{X}$, which can be interpreted as the $k$th perception of the agent, followed by cycle $k+1$ and so on. An agent is defined by a deterministic policy function $p : \mathcal{X}^* \to \mathcal{Y}^*$ with $p(x_{<k}) = y_{1:k}$ to denote output $y_{1:k} = y_1 y_2 \ldots y_k$ on input $x_{<k} = x_1 x_2 \ldots x_{k-1}$. We use the terms *policy* and *agent* interchangeably. The inputs are separated into two parts, $x_k \equiv r_k o_k$, with $r_k = r(x_k)$ representing the reward and $o_k$ representing the observation. We say $r(x_{1:m}) = \sum_{i=1}^{m} r(x_i)$ and we assume bounds on rewards with $0 \leq r_k \leq c$ for all $k$. There is uncertainty in the environment; it can be represented by a probability distribution over infinite strings, where $\mu(x_1 \ldots x_n)$ is the probability that an infinite string starts with $x_1 \ldots x_n$. In Hutter's notation [Hut04], an underlined argument $\underline{x}_k$ is a probability variable and non-underlined arguments $x_k$ represent the condition with $\mu(x_{<n}\underline{x}_n) = \mu(\underline{x}_{1:n})/\mu(\underline{x}_{<n})$. The probability that the environment reacts with $x_1 \ldots x_n$ under agent output $y_1 \ldots y_n$ is $\mu(y_1\underline{x}_1 \ldots y_n\underline{x}_n) = \mu(y\underline{x}_n)$. The environment is chronological, in that input $x_i$ only depends on $yx_{<i}y_i$. The horizon $m$ of the interaction is the number of cycles of the interaction. The *value* of agent $p$ in environment $\mu$ is the expected reward sum $V^{p,\mu}_{1:m} = \sum_{x_{1:m}} r(x_{1:m})\mu(y\underline{x}_{1:m})|_{y_{1:m}=p(x_{<m})}$. The optimal agent that maximizes value $V^{p,\mu}_{1:m}$ is $p^\mu = \arg\max_p V^{p,\mu}_{1:m}$, with value $V^{*,\mu}_{1:m} = V^{p^\mu,\mu}_{1:m}$. The optimal expected reward given a partial history $yx_{1:k}$ is $V^{p,\mu}_{1:m}(yx_{1:k})$.

It is possible to construct players (sets) $A$ and $B$ from the agent $p$ and environment $\mu$ where $A$ "interacts" (intersects) with $B$ only if agent $p$ can achieve a certain level of reward in $\mu$. This construction enables us to apply the results and proof techniques of section 5 to the cybernetic agent model. To translate $p$ and $\mu$ to $A$ and $B$, we fix two parameters: a time horizon $m$ and a difficulty threshold $\tau$. For every agent $p$, there is a player $A^p_m$, with $A^p_m = \{yx_{1:m} : y_{1:m} = p(x_{<m})\}$. There are several possible ways to construct a set $B$ from an environment $\mu$. One direct method is for every environment $\mu$, to define a player $B^\mu_{m,\tau}$, with $B^\mu_{m,\tau} = \{yx_{1:m} : r(x_{1:m}) \geq \tau, \mu(y\underline{x}_{1:m}) > 0\}$. Player $B^\mu_{m,\tau}$ represents all possible histories of $\mu$ (however unlikely) where the reward is at least $\tau$. If $A^p_m \cap B^\mu_{m,\tau} = \emptyset$, then environment $\mu$ is "too difficult" for the agent $p$; there is no interaction where the agent can receive a reward of at least $\tau$. We say the agent $p$ *interacts* with the environment $\mu$ at time horizon $m$ and difficulty $\tau$ if $A^p_m \cap B^\mu_{m,\tau} \neq \emptyset$.

*Example 2 (Peter and Magnus).* We present a cybernetic agent model interpretation of chess with reward based players Peter and Magnus (same rules as example 1). Peter, the agent $p$, has to be deterministic whereas Magnus, the environment $\mu$, has uncertainty. At cycle $k$, each action $y_k$ is Peter's move and each perception $x_k$ is Magnus' move. At ply $m$ in the chess game, Magnus returns a reward of 1 if Peter has won. In rounds where the game is unfinished or if Peter loses or draws, the reward is 0. The player (set) $A^p_m$ represents Peter's plays for $m$ rounds. The player (set) for Magnus with difficulty threshold $\tau = 1$ and $m$ plies, $B^\mu_{m,1}$ is the set of all games that Magnus loses in $m$ rounds or less. If $A_m \cap B^\mu_{m,1} = \emptyset$, then Peter cannot *interact* with Magnus at difficulty level 1; Peter can never beat Magnus at chess in $m$ rounds or less. If $A_m \cap B^\mu_{m,1} \neq \emptyset$ then Peter can beat Magnus at a game of chess in $m$ rounds or less.

Another construction of a player $D^\mu_{m,\tau}$ with respect to environment $\mu$, is $D^\mu_{m,\tau} = \{yx_{1:m} : \forall_k \ V^{*,\mu}_{1:m}(yx_{1:k})/V^{*,\mu}_{1:m} \geq \tau\}$. With this interpretation, player $D^\mu_{m,\tau}$ represents all histories where at each time $k$, $1 \leq k \leq m$, an agent can potentially achieve an expected reward of at least $\tau$ times the optimal expected reward. If $A^p_m \cap D^\mu_{m,\tau} = \emptyset$, then environment $\mu$ is "too difficult" for the agent $p$; there is no interaction where at *every* cycle $k$ the agent has the potential to receive an expected reward of at least $\tau V^{*,\mu}_{1:m}$.

## 3 Background in Algorithmic Information Theory

We denote finite binary strings by $x \in \{0,1\}^*$ and the length of strings by $l(x)$. Let the pairing function $\langle \cdot, \cdot \rangle$ be the standard one-to-one mapping from $\mathcal{N} \times \mathcal{N}$ to $\mathcal{N}$, where: $\langle x, y \rangle = x'y = 1^{l(l(x))} 0 l(x) x y$ and $l(\langle x, y \rangle) = l(y) + l(x) + 2l(l(x)) + 1$. The Kolmogorov complexity $C(x)$ is the length of the shortest binary program to compute $x$ on a universal Turing machine $\psi$, $C(x) = \min\{l(d) : \psi(d) = x\}$. The prefix-free Kolmogorov complexity, $K(x)$, restricts the universal machine $\psi$ so no halting program is a proper prefix of another halting program. For the rest of this paper, we use plain Kolmogorov complexity. Kolmogorov complexity is not finitely computable. The conditional Kolmogorov complexity of $x$ relative to $y$, $C(x|y)$, is defined as the length of a shortest program to compute $x$, using $y$ as an auxiliary input to the computation. The complexity of two strings $x$ and $y$ is denoted by $C(x, y) = C(\langle x, y \rangle)$. The conditional complexity of two strings is $C(x|y, z) = C(x|\langle y, z \rangle)$. The complexity of information in $x$ about $y$ is $I(x : y) = C(y) - C(y|x)$. The conditional mutual information is $I(x : y|z) = C(y|z) - C(y|x, z)$ and can be interpreted as the information $z$ receives about $y$ when given $x$. The complexity of a function $f : \{0,1\}^* \to \{0,1\}^*$ is $C(f) = \min\{C(p) : \forall_x \psi(p, x) = f(x)\}$. The Levin complexity is defined by $C_t(x) = \min_p\{l(p) + \log t(p, x) : \psi(p) = x\}$, with $t(p, x)$ being the number of steps taken by $\psi$ until $x$ is printed (without $\psi$ necessary halting). Levin complexity is computable. The complexity of a finite set $S$ is $C(S)$, the length of the shortest program $f$ from which the universal Turing machine $\psi$ computes a listing of the elements of $S$ and then halts. If $S = \{x_1, \ldots, x_n\}$, then $\psi(f) = \langle x_1, \langle x_2, \ldots, \langle x_{n-1}, x_n \rangle \ldots \rangle\rangle$. The conditional complexity $C(x|S)$ is the length of the shortest program from which $\psi$, given $S$ literally as auxiliary information, computes $x$. For every set $S$ containing $x$, it must be that $C(x|S) \leq \log |S| + O(1)$. The randomness deficiency is the lack of typicality of $x$ with respect to set $S$, with $\delta(x|S) = \log |S| - C(x|S)$, for $x \in S$ and $\infty$ otherwise. If $\delta(x|S)$ is small enough, then $x$ is a typical element of $S$; $x$ satisfies all simple properties that hold with high majorities of strings in $S$.

*Example 3 (Anatoly's Games).* Chess player Anatoly with function $f_A$ can be represented as a set $A$ (see example 1). Set $A$ is simple relative to $f_A$ and the maximum number of plies $m$, with $C(A|f_A, m) = O(1)$, where $O(1)$ is the length of code required to use $f_A$ and $m$ to enumerate all games $x \in A$.

The following theorem, used in section 5, shows that if a string $x$ is contained by a large number of sets of a certain complexity, then it is contained by a simpler set [VV04]. The enumerative complexity, $CE(\mathcal{F})$, is the Kolmogorov complexity of a non halting program that enumerates all the sets $F \in \mathcal{F}$. This theorem also holds for conditional complexity bounds, $C(F|y)$.

**Theorem 1 ([VV04]).** *Let $\mathcal{F}$ be a family of subsets of a set of strings $\mathcal{G}$. If $x \in \mathcal{G}$ is a member of each of $2^k$ sets $F \in \mathcal{F}$ with $C(F) \leq r$, then $x$ is a member of a set $F'$ in $\mathcal{F}$ with $C(F') \leq r - k + O(\log k + \log r + \log \log |\mathcal{G}| + CE(\mathcal{F}))$.*

## 4 Player Strategy Learning

Players $A$ and $B$ can learn information about each other's strategies from a single interaction (game) $x \in A \cap B$ or from their entire interaction set (all possible games) $A \cap B$. The *capacity* of a player $A$ is the maximum amount of information that $A$ can receive about another player through all possible interactions, i.e. their interaction set. It is equal to the log of the number of possible subsets that it can have, $\log 2^{|A|} = |A|$. We define the lack of typicality of a subset $S$ with respect to $A$ to be $\delta(S|A) = |A| - C(S|A)$, for $S \subseteq A$ and $\infty$ otherwise.

*Example 4 (Capacity).* Boris $B$ uses a range of black openings whereas Bill $B'$ uses only the Sicilian defence. So Boris has a higher capacity, $|B| \gg |B'|$, and can potentially learn more than Bill.

*Example 5 (Randomness Deficiency).* Let $A$ be the chess games played by Anatoly. Bob is a simple player $B'$, who only moves his knight back and forth. Set $S = A \cap B'$ represents all $A$'s games with $B'$. The randomness deficiency of these games, $\delta(S|A)$, is high, as $S$ is easily computable from $A$, with $C(S|A) \ll |A|$. Let $T \subseteq A$, in which $T = A \cap B$ are games played between Anatoly and Boris, who uses a range of chess strategies unknown to Anatoly. Then $\delta(T|A)$ is low and $C(T|A)$ is high.

If $A$ views every interaction in $A \cap B$, the amount of information $B$ reveals about itself is, $I(A \cap B : B|A)$, the mutual information between $B$ and $A \cap B$, given $A$. This term can be reduced to $C(A \cap B|A) - C(A \cap B|A, B) = C(A \cap B|A) + O(1)$. We define the amount of knowledge that $A$ received about $B$ from the interaction setp as:

$$R(B|A) = C(A \cap B|A). \tag{1}$$

The higher the randomness deficiency, $\delta(A \cap B|A)$, of an interaction set, $A \cap B$, with respect to player $A$, the less information, $R(B|A)$, player $A$ can receive about its opponent $B$, with

$$R(B|A) + \delta(A \cap B|A) = |A|. \tag{2}$$

Player $A$ receives the most information about its opponent when the randomness deficiency is $\delta(A \cap B|A) \approx 0$.

*Example 6.* Let Anatoly, $A$, and Bob, $B'$, be the players of example 5. Bob has a simple strategy and has a lower capacity $|B'| \ll |A|$, but he learns a lot from Anatoly, with $\delta(A \cap B'|B') \approx 0$ and $R(A|B') \approx |B'|$. Anatoly learns very little from Bob, with $R(B'|A) \approx 0$ and $\delta(A \cap B'|A) \approx |A|$.

Players can reveal information about themselves through a single interaction. The amount of information that $A$ received about $B$ from their interaction $x$ is

$$I(x : B|A) = C(x|A) - C(x|A, B). \tag{3}$$

A graphical depiction of the complexities relating to $A$, $B$, and $x$ can be seen in Fig. 2. We define the lack of typicality of an interaction $x$ with respect to both players to be

$$\delta(x|A, B) = \log |A \cap B| - C(x|A, B) \tag{4}$$

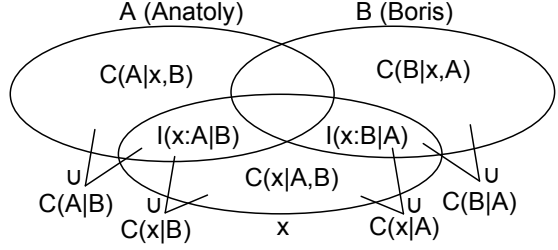

**Fig. 2.** The complexities and information of $A$, $B$, and their interaction $x$. The relationships hold up to logarithmic precision.

for $x \in A \cap B$ and $\infty$ otherwise. If $\delta(x|A,B)$ is small, then $x$ represents a typical interaction. The information passed from player $B$ to player $A$ through a single interaction is represented by

$$I(x:B|A) + \delta(x|A) = \log|A|/|A \cap B| + \delta(x|A,B). \tag{5}$$

The information passed between players through a single interaction with the same capacity is

$$I(x:B|A) + \delta(x|A) = I(x:A|B) + \delta(x|B) + O(1). \tag{6}$$

*Example 7.* Anatoly $A$ plays a game $x$ with Boris $B$ who has the same capacity with $|A| = |B|$. Anatoly tricks Boris with a King's gambit and the game $x$ follows a series of moves extremely familiar to Anatoly. Boris reacts with the most obvious move at every turn. In this case the game is simple to Anatoly, with $\delta(x|A)$ being large and $I(x:B|A)$ being small. The game is new to Boris with $\delta(x|B)$ being small and $I(x:A|B)$ being large. Thus Boris learns more than Anatoly from $x$.

If the players have a deterministic interaction, then $A \cap B = \{y\}$ and the information $A$ received from $B$ reduces to $I(y:B|A) + \delta(y|A) = \log|A|$.

## 5 Player Approximation and Interaction Complexity

We show that, given an interaction $x$ between players $A$ and $B$, $A$ can "construct" an approximate player $B'$ that has interaction $x$ using a small number of extra bits $\epsilon$, where $C(B'|A,x) = \epsilon$. We also show that the conditional complexity $C(B'|A)$ of the approximate player $B'$ is not greater than the amount of information $I(x:B|A)$ that $A$ obtains about $B$ (up to logarithmic precision). We use the simplified notation $\log A = \log|A|$. We also use the player space notation, $\mathcal{B}$, to denote a set of sets of strings.

**Theorem 2.** *Given are a player space $\mathcal{B}$ and players $A$ and $B \in \mathcal{B}$ over strings of size $n$ with $x \in A \cap B$ and $C(\mathcal{B}) = O(\log n)$. Then there is a player $B' \in \mathcal{B}$ with $x \in B'$, $C(B'|A,x) = O(s)$, and $C(B'|A) \leq I(x:B|A) + O(s)$, with $s = \log C(B|A) + \log n$.*

*Proof.* Let $r = C(B|A)$. We define $\mathcal{G}$ as the set of strings of size $n$, with $\log\log|\mathcal{G}| = \log n$. We set $\mathcal{F} = \mathcal{B}$, and so $CE(\mathcal{F}) = O(\log n)$. Let $N$ be the number of sets $S \in \mathcal{B}$, with $C(S|A) \leq r$ and $x \in S$. We first show that $C(B|A,x) \leq \log N + O(\log nr)$. There is a program, that when given $x$, $A$, $\mathcal{B}$, and $r$, with $C(\mathcal{B}, r) = O(\log nr)$, can enumerate all sets in $\mathcal{B}$ containing $x$ with conditional complexity to $A$ being less than or equal to $r$. Thus $B$ can be created using such a program and an index of size $\lceil \log N \rceil$. By the application of Theorem 1, conditional on $A$, with $k = \lfloor \log N \rfloor$, there is a set $B' \in \mathcal{F}$ with $x \in B'$ and $C(B'|A) \leq r - k + O(\log nr) \leq C(B|A) - C(B|A,x) + O(\log nr) = I(x:B|A) + O(\log nr)$. To prove $C(B'|A,x) = O(s)$, assume $B'$ is the set satisfying the above properties that minimizes $C(B'|A)$ up to precision $O(s)$. It must be that $C(B'|A,x) = O(s)$. Otherwise $C(B'|A,x) = \omega(s)$ and there is a set $B''$ satisfying properties above and $C(B''|A) \leq C(B'|A) - C(B'|A,x) + O(s) = C(B'|A) - \omega(s)$, causing a contradiction.

*Example 8 (Opponent Reconstruction).* Anatoly, $A$, plays a chess game $x$ with Boris, $B$, with $x \in A \cap B$. The players use a random string $b$ of size $C(x|A,B)$ to help decide their moves. Without using $b$, Anatoly can "construct" Bob, $B'$, an impersonator of Boris, using information from the game $x$ and $O(\log C(B|A) + \log l(x))$ bits. Bob can play the same game $x$ with Anatoly.

Given are players $A$ and $B$ who *interact*, in that $A \cap B \neq \emptyset$. We show that there exists an interacting player $B'$ that has complexity bounded by the mutual information of $A$ and $B$. If theorem 1 can be strengthened such that the enumerative complexity term $CE(\mathcal{F})$ is replaced by $CEE(\mathcal{F})$, the complexity of enumerating both the sets and the elements of the sets of $\mathcal{F}$, then the precision of theorems 3 and 4 can be strengthened with the replacement of the Levin complexity term $C_t(A)$ with Kolmogorov complexity $C(A)$.

**Theorem 3.** *Given are a player space $\mathcal{B}$ and players $A$ and $B \in \mathcal{B}$ with $A \cap B \neq \emptyset$. Then there exists a player $B' \in \mathcal{B}$ with $A \cap B' \neq \emptyset$, and $C(B') \leq I(A:B) + O(s)$, with $s = \log C(B) + \log C_t(A) + C(\mathcal{B})$.*

*Proof.* Let $r = C(B)$, $h = C_t(A)$, and $q = 2^{C(\mathcal{B})}$. We define $\mathcal{G} = \{\langle S \rangle : C_t(S) \leq r\}$, with $\langle S \rangle$ being an encoding of set $S$. This implies $\log\log|\mathcal{G}| = O(\log h)$. We define $\mathcal{F}$ with a recursive function $\lambda : \mathcal{B} \to \mathcal{F}$, with $\lambda(S) = \{\langle T \rangle \mid C_t(T) \leq h, S \cap T \neq \emptyset\}$. It must be $C(\lambda) = O(\log h)$. The enumeration complexity of $\mathcal{F}$ requires the encoding of $\mathcal{B}$ and $\lambda$, and so $CE(\mathcal{F}) = O(\log hq)$. Thus if $\langle T \rangle \in \lambda(S)$, then $T \cap S \neq \emptyset$. Let $N$ be the number of sets $S \in \mathcal{B}$, with $C(S) \leq r$ and $S \cap A \neq \emptyset$. Thus $C(B|A) \leq \log N + O(\log hqr)$, as there is a program, when given $A$, $r$, $\mathcal{B}$, and an index of size $\lceil \log N \rceil$, that can return any such $S$. By the application of Theorem 1, with $x = \langle A \rangle$ and $k = \lfloor \log N \rfloor$, there is a set $F \in \mathcal{F}$ with $x \in F$ and $C(F) \leq r - k + O(\log hqr) \leq C(B) - C(B|A) + O(\log hqr) = I(A:B) + O(\log hqr)$. A set $B' \in \mathcal{B}$, with $\lambda(B') = F$, can be easily recovered from $F$ by enumerating all sets in $\mathcal{B}$, applying $\lambda$ to each one, and selecting the first one which produces $F$. So $C(B') \leq C(F) + O(\log q) \leq I(A:B) + O(\log hqr)$. Since $\langle A \rangle \in \lambda(B')$, it must be that $A \cap B' \neq \emptyset$.

We show that if a player $A$ *interacts* with numerous players of a given complexity and uncertainty, then there exists a simple player $B'$ who interacts with $A$ with the same uncertainty.

**Theorem 4.** *Given are player space $\mathcal{B}$, player $A$ and $2^k$ players $B \in \mathcal{B}$ where for each $B$, $0 < |A \cap B| \leq c$ and $C(B) \leq r$. There is a player $B' \in \mathcal{B}$ such that $0 < |A \cap B'| \leq c$ and $C(B') \leq r - k + O(s)$, with $s = \log C_t(A) + \log c + \log k + \log r + C(\mathcal{B})$.*

*Proof.* Let $h = C_t(A)$ and $q = 2^{C(\mathcal{B})}$. We can define $\mathcal{G} \subseteq \{0,1\}^*$ as a set of strings, each encoding a set (player) $S$ whose Levin complexity is less than or equal to $h$. This implies $\log \log |\mathcal{G}| = O(\log h)$. We represent the encoding of $S$ with $\langle S \rangle$. We define $\mathcal{F}$ with a recursive function $\lambda : \mathcal{B} \to \mathcal{F}$, with $\lambda(S) = \{\langle T \rangle \mid C_t(T) \leq h, 0 < |S \cap T| \leq c\}$. Thus it must be $C(\lambda) = O(\log ch)$. The enumeration complexity of $\mathcal{F}$ requires the encoding of $c$, $h$, and $\mathcal{B}$, with $CE(\mathcal{F}) = O(\log chq)$. Thus if $\langle T \rangle \in \lambda(S)$, then player $T$ and player $S$ have a non empty intersection of size at most $c$. From the assumptions of this theorem, $\langle A \rangle$ is covered by at least $2^k$ sets $\lambda(B) \in \mathcal{F}$ of complexity $C(\lambda(B)) \leq r + O(\log chq)$. By the application of Theorem 1, with $x = \langle A \rangle$, there is a set $F \in \mathcal{F}$ with $x \in F$, $C(F) \leq r - k + O(\log(chkqr))$. A set $B' \in \mathcal{B}$, with $\lambda(B') = F$ can be recovered from $F$ by enumerating all sets in $\mathcal{B}$, applying $\lambda$ to each one, and selecting the first one which produces $F$. Therefore $C(B'|F) \leq O(\log chq)$ and so $C(B') \leq C(F) + O(\log chq) \leq r - k + O(\log(chkqr))$. Since $\langle A \rangle \in \lambda(B')$, it must be that $0 < |A \cap B'| \leq c$, thus the theorem is proven.

*Example 9.* An example application of theorem 4 is a game of the same form as example 2. Magnus, represented by set $B$, plays $2^k$ games of against $2^k$ young players $A \in \mathcal{A}$. Furthermore the players and Magnus are deterministic with for each $A \in \mathcal{A}$, $|A \cap B| = 1$. The difficulty threshold $\tau$, is set to 1, so every one of the young players beat Magnus. By theorem 4, if all players $A \in \mathcal{A}$ have complexity at most $C(A) \leq r$, then there is a simpler player $A' \in \mathcal{A}$ that will win against Magnus, with $C(A') \leq r - k + \epsilon$ (with $\epsilon$ being of logarithmic order) and $|A' \cap B| = 1$.

## 6 Future Work: Universal Interaction

Since the agents and environments of the cybernetic agent model of Section 2.2 can be translated into set representations, there is potential application of the proof techniques used in Section 5 to Artificial Universal Intelligence [Hut04], and in particular to describe properties of the AIXI model. The universal environment, $\xi$, is defined using a form of the universal prior, $\mathbf{m}(x) = \sum_{p:\psi(p)=x} 2^{-l(p)}$, representing a semimeasure (degenerate probability) over all infinite strings, with $\xi(y\underline{x}_{1:k}) = \sum_\rho 2^{-K(\rho)} \rho(y\underline{x}_{1:n})$. The universal environment $\xi$ is the weighted summation over all chronological environments $\rho$. The term $K(\rho)$ represents the *prefix free* Kolmogorov complexity of $\rho$. The AIXI model $p_m^\xi$ is the optimal agent for the environment $\xi$ with horizon $m$, in that $p_m^\xi = \arg\max_p V_{1:m}^{p,\xi}$. The sequence of self optimizing AIXI agents for each time horizon is $\{p_i^\xi\}_{i=1,2,\ldots}$. Let

$\mathcal{M}$ be a set of environments where a sequence of self-optimizing policies $\tilde{p}_m$ exists. The sequence converges to receive the optimal average for all environments with $\forall \nu \in \mathcal{M} : \frac{1}{m} V_{1:m}^{p_m,\nu} \stackrel{m \to \infty}{\longrightarrow} \frac{1}{m} V_{1:m}^{*,\nu}$. By theorem 5.29 from [Hut04], it must be that the sequence of AIXI agents is optimal for $\mathcal{M}$ with $\frac{1}{m} V_{1:m}^{p_m^\xi,\nu} \stackrel{m \to \infty}{\longrightarrow} \frac{1}{m} V_{1:m}^{*,\nu}$. We use the conversion of agents $p$ and environments $\mu$ to sets $A_m^p$ and $D_{m,\tau}^\mu$ as introduced at the end Section 2.2. The sequence of self optimizing AIXI agents, $\{p_i^\xi\}_{i=1,2,\ldots}$, is universal with regard to interaction with respect to $\mathcal{M}$. It is easy to see that for all $\tau$ and all environments $\nu \in \mathcal{M}$, there is a number $m_{\nu\tau}$ where for all $m > m_{\nu,\tau}$, $A_m^{p_m^\xi} \cap D_{m,\tau}^\mu \neq \emptyset$. This implies a set representation of agent dynamics can be used to describe further properties of the AIXI model. There is potential for a deep connection, roughly analogously to how prefix-free Kolmogorov complexity and the universal prior are related with the Coding Theorem $K(x) = -\log \mathbf{m}(x) + O(1)$ [LV08].

## 7 Conclusions

We used Algorithmic Information Theory to quantify the information exchanged between agents that interact in non-cooperative games (Sec. 4). We have shown that an agent $A$ can construct an approximation of his opponent $B$ using information from a single interaction (game) with $B$ (Th. 2). We have shown that if an agent $B$ with superfluous information interacts with an environment $A$ and achieves a certain reward, then there exists another agent $B'$ without this information that can achieve the same reward (Th. 3). We have also shown that if multiple agents interact with an environment to achieve a certain reward, then there exists a simple agent who can achieve the same reward (Th. 4). Our constructions are compatible with Universal Artificial Intelligence, in that the AIXI model can be interpreted as universal with regard to interactions with environments (Section 6).

## References

CT91. T. Cover and J. Thomas. *Elements of Information Theory.* Wiley-Interscience, New York, NY, USA, 1991.

GP07. B. Goertzel and C. Pennachin. *Artificial General Intelligence (Cognitive Technologies).* Springer-Verlag New York, Inc., Secaucus, NJ, USA, 2007.

GTV01. P. Gács, J. Tromp, and P. Vitanyi. Algorithmic Statistics. *Information Theory, IEEE Transactions on*, 47(6), 2001.

Hut04. M. Hutter. *Universal Artificial Intelligence: Sequential Decisions based on Algorithmic Probability.* Springer, Berlin, 2004.

LV08. M. Li and P. Vitányi. *An Introduction to Kolmogorov Complexity and Its Applications.* Springer Publishing Company, Incorporated, 3 edition, 2008.

RN09. S. Russell and P. Norvig. *Artificial Intelligence: A Modern Approach.* Prentice Hall, 3rd edition, 2009.

VV04. N. Vereshchagin and P. Vitányi. Algorithmic Rate Distortion Theory. arXiv:cs/0411014v3, 2004. http://arxiv.org/abs/cs.IT/0411014.

VV10. N. Vereshchagin and P. Vitányi. Rate Distortion and Denoising of Individual Data using Kolmogorov Complexity. *IEEE Transactions on Information Theory*, 56, 2010.